\documentclass{article}
\usepackage[T1]{fontenc}
\usepackage{spconf,amsmath,amssymb,graphicx,booktabs}
\title{AMemOps: Evaluating Memory Updates in Long Audio Streams}
\name{\shortstack{Peiwei Chang\textsuperscript{1}, Li Zhang\textsuperscript{2}, Mark D. Plumbley\textsuperscript{3},
Philip J.B. Jackson\textsuperscript{1}, Wenwu Wang\textsuperscript{1}}}
\address{\textsuperscript{1}Centre for Vision, Speech and Signal Processing, University of Surrey, UK\\
\textsuperscript{2}Department of Computer Science, Royal Holloway, University of London, UK\\
\textsuperscript{3}Department of Informatics, King's College London, UK
}

\begin{document}
\ninept
\maketitle
\raggedbottom
\brokenpenalty=1000
\clubpenalty=10000

\begin{abstract}
Audio memory must recognise recurring sources across acoustic changes while keeping their attributes up to date.
Overall accuracy does not reveal how these demands interact. We introduce Audio Memory Operations (AMemOps), a benchmark for latest-attribute retrieval, preservation of unchanged attributes
and absent-source rejection in long audio streams.
Paired controls vary recordings, training identities and earlier events to examine their effects on updating.
Training on other recordings of the test instruments increases a reference delta-rule memory's update advantage
over matched attention by 13.4 percentage points.
At retrieval delays of 120\,s, matched histories separate benefits of consistent repetition from costs of conflicting earlier values, while additive-memory controls identify benefits of erasure.
\end{abstract}


\section{Introduction}
Understanding long audio streams requires recognising recurring sources and tracking their changing attributes~\cite{shinn2008objects}.
When a violin returns at a different pitch and level, a system should retrieve its current level while retaining its unchanged spatial position. Such tracking could support persistent audio assistants and acoustic monitoring, where new observations can make earlier descriptions outdated.
Retrieval errors can therefore arise from failures to recognise the source, access its history or update an attribute.

Existing models offer different ways to retain a source's history. Attention stores previous observations for later retrieval, whereas delta-rule memories update key--value associations in a fixed-size state~\cite{schlag2021fastweights,yang2024deltanet}.
Recent variants refine how this state is erased~\cite{li2026eda,gupta2026qed}.
Both depend on reliable representations: attention must select the latest matching occurrence, while recurrent updates require consistent source keys~\cite{arora2025mechanistic,ghriss2026kata}.

Existing benchmarks address parts of this problem. Explicit-key tasks test retention and replacement~\cite{arora2023zoology,poli2024mad,dowling2026memory}; conversational MemOps tests whether answers reflect updated facts~\cite{hao2026memops}.
Neither requires recognising a source across acoustic recordings. Audio benchmarks cover attributes, timing and long-context retrieval~\cite{fayek2020daqa,abdelnour2023naaqa,jin2026rail,ghosh2025af2,he2025audiomarathon,vakada2026longaudiorag,gosai2026audiomultichallenge}.
EnvMem diagnoses retrieval of an earlier sound~\cite{xiao2026envmem}, while EmoUpdate addresses interference from prior emotion labels supplied as text~\cite{liu2026emoupdate}.
These studies leave open how recording variation and obsolete acoustic observations jointly affect retrieval of a source's latest attributes.

\begin{figure*}[!t]\centering\includegraphics[width=\textwidth]{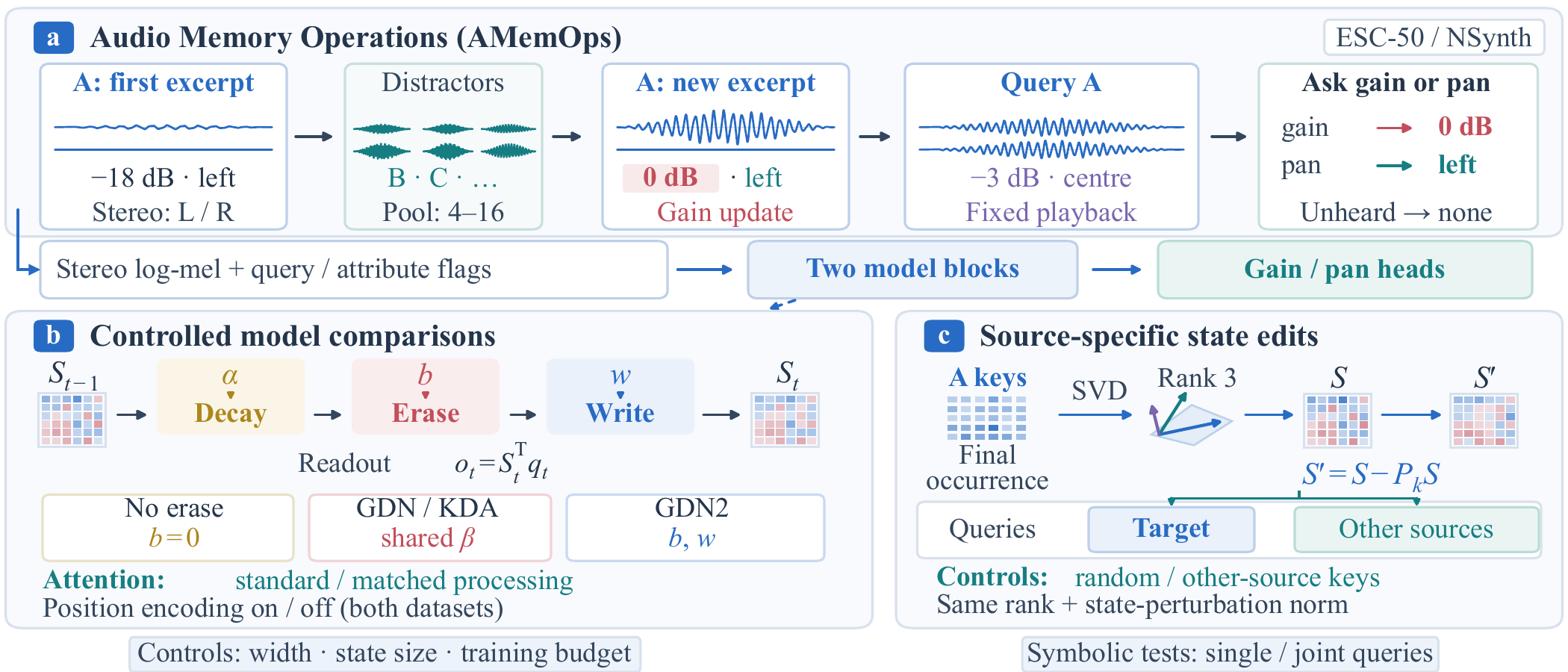}
\par
\caption{AMemOps overview. (a) Latest gain or pan queries; source labels are illustrative, not inputs. (b) Decay, erasure, writing and readout, comparing Gated DeltaNet (GDN)~\cite{yang2024gdn}, Kimi Delta Attention (KDA)~\cite{kimi2025linear}, Gated DeltaNet-2 (GDN2)~\cite{hatamizadeh2026gdn2} and attention. (c) Source-key state edits test selective access against norm-matched controls.}\label{fig:framework}\end{figure*}

We introduce the \textbf{Audio Memory Operations (AMemOps)} framework (Fig.~\ref{fig:framework}).
Our main contributions are: \textbf{(1) An audio memory benchmark:} environmental-sound and instrument streams test latest-attribute retrieval, unchanged-attribute retention and absent-source rejection without source labels. \textbf{(2) Paired diagnostics:} recording and training controls test acoustic variation and source familiarity,
while matched histories vary earlier attributes with the latest event, query and answer fixed. \textbf{(3) Controlled model comparisons:} we compare attention and recurrent memory under these conditions, using additive-memory controls and source-specific state edits to examine erasure and selective access.

\section{Methods}\label{sec:methods}
\textbf{Audio streams and queries.} Each sound class or instrument identity has independently controlled gain and stereo pan (Fig.~\ref{fig:framework}a).
Nonoverlapping excerpts, distractors and queries form 2,048-frame streams at 25\,Hz (81.92\,s).
Recordings are normalised to root mean square (RMS) amplitude 0.05 before extracting 18--32-frame excerpts (0.72--1.28\,s).
Excerpts receive 10-ms fades, uniformly sampled gain $\{-18,-12,-6,0\}$\,dB and equal-power pan $\{0,1/3,2/3,1\}$ (left to right).
Panning preserves total event energy. We then add independent white Gaussian noise per channel, 45\,dB below the RMS reference, without final normalisation or clipping.

A 12-frame query requests the source's latest gain or pan, or \emph{none} for a previously absent source.
Queries use $-3$\,dB and centre pan, outside the answer classes, to avoid revealing the answer.
Query frames update model memory but not ground-truth attributes. Inputs combine stereo log-mel features with query, gain-request and pan-request flags.
Features use 64 bands per channel (16\,kHz, 1,024-sample Hann window, 640-sample hop, 50--7,800\,Hz).
Separate gain and pan classifiers predict four values plus \emph{none}, using cross-entropy at query end.

The generator proposes 10 initial-value (no prior change), 8 post-update, 6 similar-source interference and 2 absent-source queries per stream, dropping events that do not fit.
Post-update queries request gain and pan separately; other queries select either uniformly. A retained attribute has never changed, although the other has. Proposed onset offsets between related events are log-uniform between 2 and 60\,s.
Retrieval delay runs from the latest source event's end to query end.
\begin{table}[t]\centering\small
\vspace{-10pt}
\caption{Model components: two pre-normalised blocks, width $d{=}128$. Conv-4 is four-frame depthwise causal convolution, and output normalisation is per head.
SiLU: sigmoid linear unit; RMSNorm: root mean square layer normalisation; RoPE: rotary position embeddings~\cite{su2021rope}.}\label{tab:setup}
\vspace{5pt}
\setlength{\tabcolsep}{2pt}
\begin{tabular}{@{}lccc@{}}\toprule
& \multicolumn{2}{c}{Attention} & \\
Mixer component & Standard & Matched & Recurrent\\\midrule
Heads $\times$ dimension & $4\times32$ & $2\times64$ & $2\times64$\\
Q/K/V processing & None & Conv-4 + SiLU & Conv-4 + SiLU\\
Q--K normalisation & None & $L_2$ & $L_2$\\
Output norm. and gate & None & \shortstack{RMSNorm +\\SiLU} & \shortstack{RMSNorm +\\SiLU}\\
RoPE & Optional & Optional & None\\
\bottomrule\end{tabular}
\end{table}
\begin{table}[t]\centering\small
\vspace{-10pt}
\caption{Query-weighted accuracy (\%) on 48 AMemOps streams after 12,000 (12k) and 24,000 (24k) training steps, averaged over three seeds.}\label{tab:main}
\vspace{5pt}
\begin{tabular}{lcccc}
\toprule
& \multicolumn{2}{c}{all queries} & \multicolumn{2}{c}{query type, 24k}\\
model & 12k & 24k & absent & updated\\
\midrule
\multicolumn{5}{c}{\textit{ESC-50 environmental sounds}}\\
GDN2          & \textbf{68.8} & 69.1 & \textbf{60.1} & 67.5\\
KDA           & 67.3 & 67.8 & 58.0 & 65.8\\
GDN           & 65.3 & 67.0 & 49.7 & 65.4\\
no-erase      & 63.6 & 65.3 & 49.3 & 63.2\\
Mamba-2-style & 61.4 & 65.5 & 50.7 & 60.1\\
standard + RoPE & 66.2 & 67.9 & 43.4 & 65.1\\
standard, no RoPE & --- & 66.9 & 38.9 & 62.9\\
matched + RoPE & --- & 71.9 & 55.6 & \textbf{69.6}\\
matched, no RoPE & --- & \textbf{72.7} & 59.0 & 67.0\\
\midrule
\multicolumn{5}{c}{\textit{NSynth instruments}}\\
GDN2          & 76.2 & 90.4 & 85.4 & 88.8\\
KDA           & 73.3 & 90.3 & 84.4 & 88.7\\
GDN           & \textbf{77.7} & \textbf{93.6} & \textbf{90.3} & \textbf{92.8}\\
no-erase      & 67.9 & 85.9 & 74.0 & 81.6\\
Mamba-2-style & 64.5 & 78.1 & 61.5 & 68.4\\
standard + RoPE & 56.3 & 68.8 & 53.5 & 61.7\\
standard, no RoPE & --- & 62.0 & 41.7 & 53.2\\
matched + RoPE & --- & 87.1 & 77.4 & 73.6\\
matched, no RoPE & --- & 86.4 & 80.6 & 69.5\\
\bottomrule
\end{tabular}
\end{table}

\textbf{Memory models.} Gated DeltaNet-2 (GDN2)~\cite{hatamizadeh2026gdn2} combines key-directed erasure~\cite{graves2014ntm} and recurrent linear attention~\cite{katharopoulos2020linear}.
Separate decay, erase and write gates enable controlled ablations.

\begin{samepage}
Each head updates $S_t\in\mathbb{R}^{d_k\times d_v}$ from learned query and key vectors $q_t,k_t\in\mathbb{R}^{d_k}$ and value $v_t\in\mathbb{R}^{d_v}$:
\begin{align}
S_t&=\big(I_{d_k}-k_t(b_t\odot k_t)^{\!\top}\big)\,\mathrm{Diag}(\alpha_t)\,S_{t-1}+k_t(w_t\odot v_t)^{\!\top},\label{eq:update}\\
o_t&=S_t^{\top}q_t.\label{eq:readout}
\end{align}
\end{samepage}
Decay $\alpha_t$ scales old contents, $b_t$ erases along the current key, and $w_t$ writes the new key--value association (Fig.~\ref{fig:framework}b).
Readout returns $o_t\in\mathbb{R}^{d_v}$ from the updated state.
Gates satisfy $\alpha_t,b_t\in[0,1]^{d_k}$ and $w_t\in[0,1]^{d_v}$. $I$, $\odot$ and $\mathrm{Diag}$ denote identity, elementwise product and diagonal-matrix construction.

Kimi Delta Attention (KDA)~\cite{kimi2025linear} ties $b_t{=}\beta_t\mathbf1_{d_k}$ and $w_t{=}\beta_t\mathbf1_{d_v}$ with scalar $\beta_t$.
Scalar decay yields Gated DeltaNet (GDN)~\cite{yang2024gdn}. \emph{No-erase} sets $b_t{=}0$ with channel-wise writing and decay~\cite{yang2024gla}. \emph{Mamba-2-style} also sets $b_t{=}0$, fixes $w_t{=}\mathbf1_{d_v}$ and uses scalar decay~\cite{dao2024mamba2}. \emph{Matched attention} uses causal softmax with the recurrent models' head dimensions and input/output processing (Table~\ref{tab:setup}).

\textbf{Recording controls.} Fixed recordings replay each source's first recording from the start in events and queries, preserving timing, attributes, answers and noise.
Intermediate controls cyclically reuse two or four recordings per source; an unrestricted control uses all recordings, also starting at zero.
\begin{table}[t]\centering\small
\vspace{-10pt}
\caption{Mean per-stream NSynth accuracy (\%) on original (O) and fixed (F) recordings: 192 held-out streams per set, 24k models from Table~\ref{tab:main}, three seeds.}\label{tab:fixed}
\vspace{5pt}
\setlength{\tabcolsep}{2.1pt}
\begin{tabular}{@{}lrrrrrr@{}}\toprule
& \multicolumn{2}{c}{Initial} & \multicolumn{2}{c}{Updated} & \multicolumn{2}{c}{Retained}\\
\cmidrule(lr){2-3}\cmidrule(lr){4-5}\cmidrule(l){6-7}
Model & O & F & O & F & O & F\\\midrule
\multicolumn{7}{c}{\textit{Seen identities (Set 1)}}\\
GDN2 & 92.2 & 99.4 & 89.2 & 98.8 & 94.1 & 99.5\\
GDN & 93.6 & 98.7 & 92.0 & 98.8 & 95.6 & 99.2\\
no-erase & 88.9 & 98.6 & 80.9 & 95.3 & 92.7 & 99.1\\
matched + RoPE & 91.8 & 99.6 & 77.1 & 98.7 & 94.9 & 99.7\\
matched, no RoPE & 91.8 & 99.7 & 72.2 & 98.1 & 94.5 & 99.4\\\midrule
\multicolumn{7}{c}{\textit{Unseen identities (Set 2)}}\\
GDN2 & 43.9 & 88.4 & 39.4 & 87.6 & 53.3 & 92.6\\
GDN & 35.4 & 69.9 & 34.3 & 72.8 & 43.5 & 80.3\\
no-erase & 44.4 & 82.3 & 40.2 & 76.2 & 54.8 & 90.1\\
matched + RoPE & 45.9 & 94.6 & 40.9 & 92.6 & 58.5 & 95.3\\
matched, no RoPE & 43.2 & 93.9 & 39.2 & 92.1 & 56.6 & 94.9\\
\bottomrule\end{tabular}
\end{table}

\textbf{Matched event histories.} We vary the earlier target event while fixing the latest event, query and answer.
The earlier target matches both latest attributes (no update), or differs only in the queried (updated) or other attribute (retained).
We evaluate the 24k models in Table~\ref{tab:main} on NSynth instrument identities seen during training.
All operation, delay, load and recording conditions share 192 single-query templates: two repeats per queried attribute, latest gain--pan pair and possible value change.

Two 0.96-s target events start 4\,s apart, the latest ending at 20\,s. Queries end at 40 or 140\,s in 4,096-frame inputs (163.84\,s): delays of 20 or 120\,s, within or beyond the training context.

Sixteen distinct distractor sources appear once at evenly spaced onsets between target and query.
Low load retains events 1, 6, 11 and 16. Longer delays stretch spacing, keeping counts, identities and durations fixed.
Noise is shared. At 120\,s, original-recording controls for GDN2, no-erase and no-RoPE attention omit the earlier target or replay the first distractor there, preserving later audio.

\textbf{State interventions.} Source-key projections test selective access (Fig.~\ref{fig:framework}c).
Per-head $P_k$ projects onto the top three right singular vectors of target keys at the final occurrence.
We apply $S\leftarrow S-P_kS$ in both layers. Eligible queries concern target or other-source bindings established before the edit and not refreshed before querying.
Controls use three random subspaces or the most overlapping other-source key subspace.
Per-layer rescaling over both heads matches the target edit's Frobenius norm (relative error $<10^{-6}$).

\section{Experiments}\label{sec:experiments}
\textbf{Data.} ESC-50~\cite{piczak2015esc} supplies 50 classes: folds 1--4 for training and fold 5 for evaluation.
The NSynth models in Tables~\ref{tab:main}--\ref{tab:fixed} train on official validation notes and evaluate on test notes~\cite{engel2017nsynth}.
Set~1 comprises 53 instruments shared across these splits at MIDI pitches 36--84, with disjoint note IDs (8,096 training, 2,632 evaluation).
The generator varies ESC-50 excerpt starts and NSynth recordings.
Mean source spectra pooled across training and evaluation recordings guide stream generation: similar pairs use the upper correlation quartile, dissimilar pairs the lower.

\textbf{Architecture and training.} Two recurrent layers each contain two $64\times64$ states, totalling 64\,KiB at 32-bit precision.
Matched attention's learned query--key scales~\cite{henry2020qknorm} start at $\sqrt{d_k}=8$.
Log-spaced decay rates $-\log\alpha\in[10^{-3},10^{-1}]$ initialise multiple timescales; erase and write gates start at $\operatorname{sigmoid}(1)$.
AdamW uses batch size 40, peak learning rate $10^{-3}$, 300-step warm-up and cosine decay.
We report final 12k and 24k checkpoints from three seeds. Training varies recordings, similarity and distractor counts (4, 8 or 16). Attention has 450k parameters; GDN2 has 549k. Symbolic controls use unit-vector identities, two 16-value attributes plus \emph{none}, one layer, one head, two seeds and 20k steps.

\begin{figure}[t]
\centering\includegraphics[width=\columnwidth]{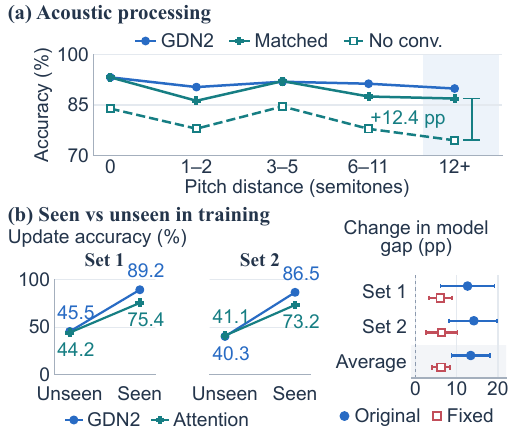}
\par
\caption{NSynth (24k, three seeds). (a) Pitch distance from query to latest same-source event.
Dashed: matched attention without convolution (9--48 streams/bin, at least two queries/stream). (b) Left: update accuracy on identical test streams for identities seen/unseen in training.
Right: change in the GDN2--attention gap (Eq.~\ref{eq:exposure}). Attention uses RoPE; bars show 95\% CIs.}\label{fig:familiarity}
\end{figure}
\begin{figure}[t]
\centering\includegraphics[width=\columnwidth]{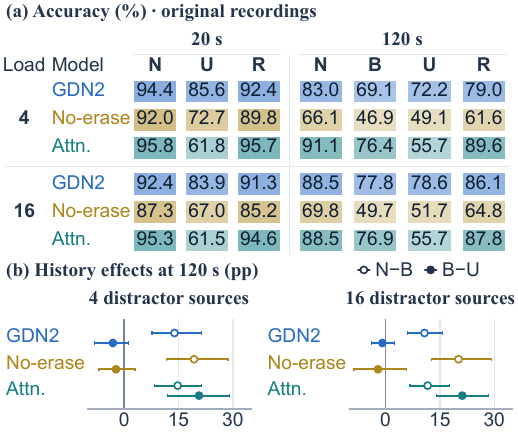}
\par
\caption{NSynth Set~1, Table~\ref{tab:main} models at 24k, three seeds. (a) Accuracy: consistent (N), earlier target absent (B), updated (U), retained (R).
Load counts distractor sources. (b) Repetition benefit N$-$B and net conflict cost B$-$U, with marginal 95\% CIs.
Each condition has 192 paired queries. Attention has no RoPE.}\label{fig:history}
\end{figure}

\textbf{Identity controls.} We test frozen models on two disjoint sets of 53 unseen instruments (Sets~2 and~3), matched by family, source type and pitch--velocity coverage.
Each has 2,632 notes and 192 streams. Calibration recordings with disjoint pitches define similarity pools.

\textbf{Training familiarity.} We train GDN2 and matched RoPE attention separately on Sets~1 and~2, then test on both, keeping each set's 192 test streams fixed.
An instrument is \emph{seen} when training includes other recordings of it, otherwise \emph{unseen}.

These models (Fig.~\ref{fig:familiarity}b) use 6,454 training, 1,589 calibration and 2,632 evaluation recordings per set, excluding shared note IDs and exact training--evaluation waveform duplicates.
Calibration recordings define similarity pools. Training uses 24k steps and three seeds per training set (12 runs).

\begin{samepage}
Let $g_{c,s}$ be the GDN2--attention update accuracy gap on test set $c$ under training status $s$. Its mean change with familiarity is:
\begin{equation}
I=\tfrac12\sum_{c=1}^{2}\big(g_{c,\mathrm{seen}}-g_{c,\mathrm{unseen}}\big).\label{eq:exposure}
\end{equation}
\end{samepage}

\textbf{Statistics.} Paired analyses average eligible gain and pan queries within streams, then weight non-empty streams and seeds equally.
Attribute-specific scores repeat this averaging on their own queries. Differences are in percentage points (pp), with 95\% confidence intervals (CIs).
In 100,000 bootstrap resamples, seeds are drawn independently across models and training sets; streams are paired across models and conditions but independent across test sets.
Four primary history contrasts test delay and load effects on GDN2's update gaps against no-erase and no-RoPE attention, averaging over the other factor on original recordings.
Simultaneous intervals use the 95th percentile of the maximum absolute centred bootstrap error.
\begin{figure}[t]\centering
\includegraphics[width=\columnwidth]{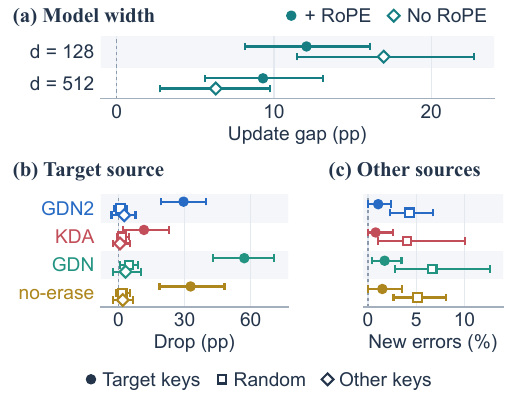}
\par
\caption{(a) GDN2's update advantage over matched attention (NSynth, 24k, 929 queries in 192 streams). (b,c) Rank-3 edits (ESC-50, 12k):
52 target and 148 other-source queries in 26 streams before target reappearance. (b) Net target accuracy loss. (c) Correct-to-wrong transitions divided by all eligible other-source queries.
Stream-weighted means, three seeds, 95\% CIs.}\label{fig:capacity_access}
\end{figure}

\subsection{Acoustic processing and recording variation}\label{sec:recognition}
\textbf{Input processing.} GDN leads on instruments, while matched attention leads on environmental sounds (Table~\ref{tab:main}).
Matching recurrent heads and processing improves NSynth attention accuracy by 21.4\,pp $[19.0,23.7]$, averaged over RoPE settings.
Convolution helps retrieval across large pitch differences (Fig.~\ref{fig:familiarity}a).

\textbf{Recording reuse improves retrieval.} Fixed recordings improve all models and close update gaps on seen instruments (Table~\ref{tab:fixed}).
With original recordings, GDN2 leads matched attention on updates by 12.1 and 17.0\,pp with and without RoPE.
Retained-attribute differences are $-0.8$ and $-0.4$\,pp (CIs include zero). Fixing recordings improves matched RoPE attention's gain and pan update accuracy by 20.6 and 20.3\,pp, respectively.
With two or four recordings per source, the gap remains 12--13\,pp; with one, both models approach perfect accuracy.
On original recordings, excluding queries whose own or earlier same-source recordings match training audio (same instrument and pitch; first-second waveform cosine $>0.9999$) retains 12.0 and 16.1\,pp advantages (616 updates, 185 streams).

\subsection{Seen and unseen instruments}\label{sec:exposure}
On unseen Set~2 instruments, GDN2's update gaps against matched attention are $-1.5$\,pp with RoPE and $+0.2$\,pp without it (CIs include zero).
The gap against RoPE attention is $+1.8$\,pp $[-2.3,5.9]$ on Set~3. Initial and retained accuracy also fall (Table~\ref{tab:fixed}), showing difficulty beyond updates.
Recording reuse improves RoPE attention's pan update accuracy by 50.9--54.1\,pp across the unseen sets.

Training on other recordings of the same test instruments increases GDN2's update advantage over matched RoPE attention by $I=13.4$\,pp $[8.7,18.0]$ (Fig.~\ref{fig:familiarity}b): 9.6\,pp $[3.8,15.3]$ for gain and 16.8\,pp $[11.3,22.4]$ for pan.
Fixed recordings reduce $I$ to 6.2\,pp $[4.0,8.4]$, a paired reduction of 7.2\,pp $[3.1,11.4]$.

\subsection{Conflicting history at long retrieval delays}\label{sec:history}
At 120\,s with four distractors, matched no-RoPE attention leads GDN2 without updates but trails after updating (Fig.~\ref{fig:history}).
Across delays and loads, its update loss relative to no update is 33--35\,pp, versus 9--11\,pp for GDN2.
Retained-attribute losses are only 0--5\,pp.

At 120\,s, attention accuracy without the earlier target is 76.4\% with four distractors and 76.9\% with sixteen.

\begin{samepage}
Against these baselines, conflicting history reduces accuracy by 20.7\,pp $[12.0,29.0]$ and 21.2\,pp $[14.1,28.5]$, while consistent history adds 14.8 and 11.6\,pp.
Replacing the earlier target with a distractor gives similar costs (20.1 and 21.0\,pp).
GDN2 and no-erase show no clear net loss (CIs include zero). \par
\end{samepage}

Previous values account for 78\% of no-RoPE attention's update errors and 23\% of GDN2's at 120\,s with four distractors (pooled seeds).
The four pre-specified interactions show no clear increase in GDN2's update advantage with delay or load.
RoPE attention falls below 3\% on all operations at 120\,s.

\subsection{Erasure, capacity and selective access}
\textbf{Erasure improves updating.} With fixed recordings, GDN2 exceeds no-erase on updates by 11.3\,pp $[1.9,24.8]$ on Set~2 and 9.5\,pp $[0.3,20.4]$ on Set~3.
Excluding Set~3 streams with duplicate evaluation notes leaves a 9.0\,pp gap $[-0.2,19.9]$ (815 updates in 172 streams).
The advantage is 5.5\,pp $[3.4,7.7]$ on 192 ESC-50 streams with repeated updates (10 update sources, up to four updates each).
Symbolic single and joint queries likewise yield 82.7--85.1\% per-attribute update accuracy for GDN2, KDA and GDN, versus 58.2\% for no-erase.

\textbf{Width control.} Widening $d$ from 128 to 512 with the same two 64-dimensional heads preserves GDN2's update advantage: 9.3\,pp $[5.6,13.1]$ with RoPE and 6.3\,pp $[2.7,9.7]$ without it (Fig.~\ref{fig:capacity_access}a).

\textbf{Selective access.} Source-key edits cause larger target losses and fewer other-source errors than random edits in erasing and additive memories (Fig.~\ref{fig:capacity_access}b,c).

\section{Conclusion}
AMemOps evaluates retrieval of current attributes and preservation of unchanged ones.
Its paired controls show how acoustic variation and familiarity shape model comparisons, and distinguish benefits of consistent repetition from costs of conflicting observations.
Additive controls show when erasure improves updating.
\clearpage

\section{Acknowledgements}\label{sec:acknowledgements}

This work was supported by the UKRI AI Centre for Doctoral Training
in AI for Digital Media Inclusion [grant number EP/Y030915/1]
and by the Engineering and Physical Sciences Research Council (EPSRC)
[grant number EP/Y028805/1].
For the purpose of open access, the authors have applied a Creative Commons
Attribution (CC BY) licence to any Author Accepted Manuscript version
arising from this submission.

\end{document}